\documentclass[11pt,a4paper]{article}
\pdfoutput=1
\usepackage[UKenglish,english]{babel}
\usepackage{jcappub}
\usepackage{subfig}
\usepackage{lpic}
\usepackage[abs]{overpic}

\newcommand{\be}{\begin{equation}}
\newcommand{\ee}{\end{equation}}
\newcommand{\beq}{\begin{equation}}
\newcommand{\eeq}{\end{equation}}
\newcommand{\beqa}{\begin{align}}
\newcommand{\eeqa}{\end{align}}

\newcommand{\abs}[1]{\ensuremath{\left\lvert #1 \right\lvert}}

\newcommand{\dbar}{{\mathchar'26\mkern-11mu\mathrm{d}}}

\renewcommand{\geq}{\geqslant}
\renewcommand{\leq}{\leqslant}

\newcommand{\ds}{\mathrm{d}}

\newcommand{\ssl}{\mathrm{s}}

\newcommand{\Ical}{\ensuremath{\mathcal{I}}}

\newcommand{\Ocal}{\ensuremath{\mathcal{O}}}
\newcommand{\Pcal}{\ensuremath{\mathcal{P}}}

\newcommand{\Ecal}{\ensuremath{\mathcal{E}}}

\newcommand{\intfb}[1]{\ensuremath{\int\dbar^4k \,}}
\newcommand{\inttb}[1]{\ensuremath{\int\dbar^3k \,}}

\newcommand{\dbd}[2]{\ensuremath{\frac{\ds #1}{\ds #2}}}

\newcommand{\Mpl}{M_\text{Pl}}

\begin{document}

\selectlanguage{UKenglish}

\title{G-Bounce \date{\today}}
\author[a]{Damien A. Easson,}
\author[b]{Ignacy Sawicki}
\author[c]{and Alexander Vikman}

\affiliation[a]{Department of Physics  \& School of Earth and Space Exploration  \& Beyond Center,\\
Arizona State University, Tempe, AZ, 85287-1504, USA}
\affiliation[b]{Institut f\"ur Theoretische Physik, Ruprecht-Karls-Universit\"at Heidelberg,
\\ Philosophenweg 16, 69120 Heidelberg, Germany}
\affiliation[c]{Theory Division,  CERN,CH-1211 Geneva 23, Switzerland}

\emailAdd{easson@asu.edu}
\emailAdd{ignacy.sawicki@uni-heidelberg.de}
\emailAdd{alexander.vikman@cern.ch}


\abstract{We present a wide class of models which realise a bounce in a spatially flat Friedmann universe in standard General Relativity. The key ingredient of the theories we consider is a noncanonical, minimally coupled scalar field belonging to the class of theories with Kinetic Gravity Braiding / Galileon-like self-couplings. In these models, the universe smoothly evolves from contraction to expansion, suffering neither from ghosts nor gradient instabilities around the turning point. The end-point of the evolution can be a standard radiation-domination era or an inflationary phase. We formulate necessary restrictions for Lagrangians needed to obtain a healthy bounce and illustrate our results with phase portraits for simple systems including the recently proposed Galilean Genesis scenario.}
\subheader{CERN-PH-TH/2011-203}

\maketitle

\section{Introduction}

Was there a beginning of time? Or in more technical words: was there a spacelike boundary of the quasiclassical spacetime, beyond which the universe was necessarily in a strong quantum gravity regime? If there was a beginning, was the universe collapsing or expanding immediately afterwards? Was the universe born infinitely (Planckian) small or infinitely large? 
If the universe experienced an early period of inflation, as all observations currently suggest, what happened before inflation \cite{Guth:1980zm,Linde:1981mu}. All these basic questions are of fundamental importance and remain interesting even if one disregards the possible observational consequences. 

Asking these questions has already led to unexpected discoveries.  Indeed the first cosmological model with a quasi-de-Sitter stage \cite{Starobinsky:1980te} and the cosmological perturbations within it \cite{Mukhanov:1981xt}\footnote{See \cite{Mukhanov:2003xw} for the English translation.} was invented as an attempt to explain how the universe could have avoided the initial singularity. Twenty years later it was nonetheless proven that inflation with canonical kinetic terms did not solve the singularity  problem \cite{Borde:2001nh}. 

Bouncing cosmological models provide an interesting possible alternative to standard Big-Bang cosmology e.g. \cite{Tolman:1931zz, Veneziano:1991ek, Mukhanov:1991zn,Gasperini:1992em, Brandenberger:1993ef,Brustein:1997cv, Khoury:2001wf, Khoury:2001bz, Gasperini:2002bn}.\footnote{For some recent reviews on bouncing cosmologies see \cite{Novello:2008ra, Lehners:2008vx, Brandenberger:2011gk, Lehners:2011kr}.}
However, such models are plagued by significant obstacles and frequently exhibit pathological behavior, for criticism see e.g.\ \cite{Linde:2007fr}. Many of these pathologies result from the need \cite{Hawking:1969sw,MolinaParis:1998tx,Cattoen:2005dx} for the energy-momentum tensor to violate the null energy condition (``NEC'': $T_{\mu\nu} n^\mu n^\nu \geq 0$, for all null vectors $n^\mu$) to bounce a spatially flat Friedmann universe.\footnote{Note that it is not difficult to construct a non-singular bouncing universe with positive spatial curvature which would compensate the positive energy density at the bounce. For that one does not need to violate the NEC, but the universe will be substantially closed. A classic example of such a non-singular universe is given in  \cite{Starobinsky1978}, where one can also find a slow-roll regime in an $m^2 \phi^2$ potential in a \emph{contracting universe}. For the most recent and simple example, see \cite{Graham:2011nb}.} The question of whether a stable violation of the NEC is possible is also crucial for the understanding of the current accelerated expansion of the universe, see e.g.\ \cite{Larson:2010gs,Caldwell:1999ew,Alam:2003fg,Feng:2004ad}. It has proven extremely difficult to construct local field-theoretic models violating the NEC in the context of standard general relativity. Until recently it was the common wisdom that any violation of the NEC leads to internal inconsistencies such as: ghost degrees of freedom and gradient instabilities (i.e.\ imaginary speed of sound)  \cite{Liu:2002yd, Buniy:2005vh, Dubovsky:2005xd, Caldwell:2005ai, Buniy:2006xf, Vikman:2004dc, Hu:2004kh, Babichev:2007dw,Xia:2007km}.\footnote{However, see  \cite{Onemli:2002hr, Onemli:2004mb, Kahya:2006hc, Kahya:2009sz}.   Moreover, it is unclear whether ghost-like instabilities are present in scalar theories with constraints such as \cite{Lim:2010yk}. These theories can violate the NEC without any gradient instabilities  \cite{Lim:2010yk} and can realise an oscillating nonsingular universe \cite{Cai:2010zma}.}  One may try to avoid these problems in the context of effective field theory by adding higher derivatives to the scalar-field action see e.g.\ \cite{Creminelli:2006xe, Creminelli:2007aq, Cheung:2007st, Creminelli:2008wc}. This approach is useful for a better systematic understanding of the perturbative expansion but cannot elucidate the behavior of the cosmological background. Inclusion of  generic higher derivative terms results in new ghost degrees of freedom \cite{Ostrogradsky},\footnote{For a modern and detailed discussion see \cite{Woodard:2006nt}.} which, in some cases, may be moved above the naive cut off. This procedure is rather delicate \cite{Aref'eva:2006xy, Kallosh:2007ad, Weinberg:2008hq} and implicitly incorporates the assumption that a healthy UV completion which takes care of the ghosts and gradient instabilities exists, which may not be the case \cite{Adams:2006sv}.     
If dangerous ghost-like instabilities are ignored, i.e.\ one assumes that some unknown mechanism exists which would cure them, it is easy to realize a bouncing scenario, for example, in \cite{Cai:2007qw,Cai:2007zv,Cai:2008qw}. 

Recently the situation has changed with the rediscovery of scalar-field theories with higher derivatives in the action, but which maintain second order equations of motion.  These higher-derivative theories possess only the  three standard degrees of freedom---the graviton and the scalar---allowing the theories to circumvent the conclusions of \cite{Ostrogradsky}. Originally these second-order theories were derived in \cite{Horndeski}.\footnote{A subclass of these models was  also considered later in a different context in \cite{Fairlie:1991qe,Fairlie:1992nb}. The result of \cite{Horndeski} was independently rediscovered in \cite{Deffayet:2011gz}. The equivalence of these results was shown in \cite{Kobayashi:2011nu}. The original Horndeski's theory was recalled for the first time in modern literature---``resurrected''---in \cite{Charmousis:2011bf}.} The simplest versions of these actions arise in certain modifications of gravity \cite{Dvali:2000hr} when considered in the decoupling limit \cite{lpr,Nicolis:2004qq,Gabadadze:2006tf} and were then generalized to the so-called \emph{Galileons} in \cite{Nicolis:2008in} for the fixed Minkowski metric and in \cite{Deffayet:2009wt,Deffayet:2009mn} for dynamical spacetime.\footnote{Note that general relativity does not allow the theory to maintain the Galilean symmetry in curved spacetime in a manifestly self-consistent fashion. For a reduced notion of Galileon symmetry in curved spacetime, see \cite{Germani:2011bc}.}   It is the presence of higher derivatives in the action and the corresponding kinetic mixing between the scalar and the metric which allows for a stable violation of the NEC \cite{Nicolis:2009qm, Deffayet:2010qz}.  The simplest class of these second-order theories which are minimally coupled i.e.\  which do not involve any direct couplings to the Riemann tensor, but still possess this kinetic mixing/braiding was introduced in  \cite{Deffayet:2010qz} under the name of \emph{Kinetic Gravity Braiding}.\footnote{See also Ref. \cite{Kobayashi:2010cm} where this class of models was studied slightly later under the name of \emph{G-inflation}.} This class of theories is also singled out from the most general second-order theories by the correspondence with the hydrodynamics of imperfect fluids \cite{Pujolas:2011he}. Indeed it is this imperfection which allows these theories  to avoid the pathologies pointed out in \cite{Dubovsky:2005xd} in the case of perfect fluids. 

In the current paper we will use exactly this class of theories with \emph{Kinetic Gravity Braiding} to study bouncing cosmologies. The possibility of a healthy bounce in a particular model of this class, the so-called \emph{Conformal Galileon},  was mentioned in \cite{Creminelli:2010ba} where the authors concentrated on an always-expanding and superaccelerating stage of the evolution of the universe: \emph{Galilean Genesis}. The details of this model were further investigated in  \cite{Levasseur:2011mw} and \cite{Qiu:2011cy}, with the latter work focusing on the bouncing solutions of the \emph{Conformal Galileon}.  

In this paper, we demonstrate that manifestly stable bouncing solutions in generic theories with \emph{Kinetic Gravity Braiding} are rather common. For simplicity, we concentrate on two broad categories of models: In one, the shift-symmetric scalar-field evolves in the presence of external hydrodynamical matter. In the other, the scalar field is not shift symmetric but it is the only source of energy density in the universe. For both categories we derive the generic conditions for high-frequency stability around the bounce point.  We have found that it is not difficult to avoid ghosts and gradient instabilities around the bounce. However, we would like to stress that we have not studied the issues related to a possible strong coupling of the scalar perturbations. To illustrate our general analysis we study the phase portraits of particular systems. 

We find that in most of the considered models, even though the bounces are healthy, the trajectories are not free of problems. Typically, pressure singularities \cite{Barrow:2004xh} or Big Rips \cite{Caldwell:2003vq} \footnote{For an earlier and more detailed discussion see \cite{Starobinsky:1999yw}.} are present at one of the ends of the trajectories, either in the past or in the future with respect to the bounce. At best, the trajectories evolve to or from regions of phase space where the sound speed of the perturbations becomes imaginary. This means that we cannot really trust the background dynamics which we calculate for the whole evolutionary history. As a result, our bouncing models, as they stand, do not resolve the initial cosmological singularity. Moreover, the amount of expansion or contraction between any such singularity and the bounce is rather limited. We have not touched on the generation of cosmological perturbations here, but it is clear that the mechanism would have to be different  from the inflationary one.    

However, we have found a category of models (the \emph{hot G-bounce}) which bounce and then evolve to a healthy and stable future, where the scalar has redshifted away and any other accompanying fluid present in the universe dominates the dynamics. As such, in these models we have a bounce followed by a hot Big-Bang, or possibly an inflationary period. In all such trajectories, at some point \emph{before} the bounce, the sound speed is imaginary. On the other hand, the physical energy scale at which this occurs can be made much smaller than the Planck energy. One could hope that this problem could be resolved without recourse to quantum gravity, but by modifying the scalar model in some way. We would like to note that the presence of stable bounces is so generic that we suspect that given sufficient effort one should be able to construct models which remain under control over the whole history, providing a never-singular evolution for the early universe. 

 The remainder of this paper is organized as follows: We begin in \S\ref{s:geneq} by introducing the model and discussing the general properties of a braided scalar and its perturbations. In \S\ref{sec:siffsymm}, we discuss bounces in the presence of external matter in theories which are shift symmetric. This simplifies the phase space considerably and allows us to find a prescription for models which bounce in a healthy fashion. To illustrate this class, we introduce the \emph{hot G-bounce} model in \S\ref{sec:hotgbounce}, which evolves to a radiation-domination era following the bounce. We present a selection of other models in this class in \S\ref{sec:othershift}. In \S\ref{s:nomatter}, we present the general properties required to build a successful bouncing model with negligible external matter and then discuss in detail the dynamics of the bounces in the \emph{conformal Galileon} model in \S\ref{sec:cb}. We conclude in \S\ref{sec:con}.

\section{General Properties}\label{s:geneq}

In order to realise a bounce in a spatially flat Friedmann universe, the theory has to violate the null energy condition (``NEC'') \cite{Hawking:1969sw}. The simplest system capable of exhibiting a large violation of the NEC without any linear instabilities is a \textit{kinetically braided}
(or \textit{galileon}) scalar field, which we have denoted as $\phi$. 

In order to aid the reader, we recap the main equations describing background evolution in cosmological models with \emph{Kinetic Gravity Braiding} and simple \emph{Galileons}. In addition, we provide the formulae determining the high-frequency stability of the model, i.e.\ inequalities which need to be satisfied to prevent ghost and gradient instabilities. We present the results in the form of  \cite{Deffayet:2010qz} and \cite{Pujolas:2011he}. We will assume that a spatially flat Friedmann universe is filled with the the scalar field $\phi$ and some hydrodynamical matter with energy density $\rho$ and pressure $p=w\rho$.  

The gravitational part of the action is given by the standard Einstein-Hilbert term
while the action for the scalar is\footnote{Unless explicitly stated otherwise, we use reduced Planck units where $M_{\text{Pl}}=\left(8\pi G_{\text{N}}\right)^{-1/2}=1$ and the metric signature convention $\left(+-\,-\,-\right)$.}
\be
S_{\phi}=\int\mbox{d}^{4}x\sqrt{-g}\left[K\left(\phi,X\right)+G\left(\phi,X\right) \Box \phi \right] \ ,
\label{e:action}
\ee
where
\be
X\equiv\frac{1}{2}g^{\mu\nu}\nabla_{\mu}\phi\nabla_{\nu}\phi\,,
\ee
and $\nabla_{\mu}$ denotes a covariant derivative, so that $\Box\equiv g^{\mu\nu}\nabla_{\mu}\nabla_{\nu}$.
Further it is convenient to introduce the \emph{diffusivity}, $\kappa$, which measures the deviation of the energy-momentum tensor from the perfect-fluid form:
\be
\kappa\equiv 2XG_{,X}\,.
\ee
Here and throughout the paper we use the notation $( \ )_{,X}=\partial ( \ )/ \partial X$. 
In \cite{Pujolas:2011he}, it was shown that $\dot \phi$ plays the role of an effective mass or chemical potential. We will use notation
\be\label{m}
 	m=\dot \phi \,,
\ee
henceforth. The total pressure of the scalar field in the reference frame moving with the scalar, with velocity $u_\mu \equiv \partial_\mu\phi /m$, is 
\be
\mathcal{P}=K-m^{2} G_{,\phi}-\kappa\dot{m}\,. \label{TotalPressure}
\ee
In a Friedmann universe with the Hubble parameter $H$, the shift charge density is given by
\begin{equation}
 n = K_{,m} - 2mG_{,\phi} + 3H\kappa \,. \label{e:n}
\end{equation}
The total energy density is given by an analogue of the thermodynamical Euler relation  
 \be
 \label{total energy}
 \Ecal =mn-\mathcal{P}-\kappa\dot{m}= m\left(K_{,m}-m G_{,\phi} \right)-K+3H m \kappa\,.
 \ee
The first Friedmann equation reads 
\begin{equation}
 \label{e:H2}
    H^2 =\frac{1}{3}\left( \Ecal +\rho \right)\,,
\end{equation}
while for the second Friedmann equation we have 
\begin{equation}
 \label{e:Hdot}
    \dot{H} =-\frac{1}{2} \left(\Ecal +\rho+\Pcal+p\right) =\frac{1}{2}\left( \kappa \dot m  - n m -\left(\rho+p\right)\right) \,. 
\end{equation}
The equation of motion for the scalar field can be written in the form
\begin{equation}
  D\dot m + 3n\left(H-\frac{1}{2}\kappa m\right) +\Ecal _{,\phi} = \frac{3}{2}\kappa(\rho+p)\,, \label{e:ddphi}
\end{equation}
where the positivity of the quantity $D$ implies that the perturbations of $\phi$ are not ghosts,
\begin{equation}
  D = n_{,m} + \kappa_{,\phi}+ \frac{3}{2}\kappa^2>0 \,. \label{e:D} 
\end{equation}
In the above, we have taken the Hubble parameter $H$ as an independent variable in the differentiation, $n=n\left(\phi,m,H\right)$ and  $\Ecal=\Ecal\left(\phi,m,H\right)$. Finally the absence of gradient instabilities requires the speed of propagation of acoustic perturbations of $\phi$ to be real:
\begin{equation}\label{generalSound}
c_{\ssl}^{2}=\frac{n+\dot{\kappa}+\kappa\left(H-\kappa m/2\right)}{Dm} =\frac{n+\dot{\kappa}+\kappa\left(H-\kappa m/2\right)}{\Ecal_{,m}-3\kappa\left(H-\kappa m/2\right)} >0\,.
\end{equation}

The background dynamics is described by four first-order equations:  $m=\dot \phi$, the continuity equation for matter $\dot \rho +3H\left(\rho+p \right)=0$, Eq.~\eqref{e:Hdot},  Eq.~\eqref{e:ddphi}, plus a constraint that is the first Friedmann equation \eqref{e:H2}. The system moves on a 3d hypersurface in the phase space $\left(\phi,m, \rho, H\right)$. There are two important cases in which the dynamics greatly simplify: 
\begin{itemize}
\item the scalar-field Lagrangian is symmetric with respect to constant shifts in field space: $\phi \rightarrow \phi +c$
\item one can neglect external matter: $\rho=0$
\end{itemize}
In both these cases, the system moves on a 2d surface. In the following analysis, we will concentrate on these two cases in turn.

\section{The Bounce in Shift-Symmetric Theories with External Matter}\label{sec:siffsymm}

Here, for simplicity, we are going to make an assumption that $p=w\rho$ with $w=\text{const}$. The phase space for this dynamical system is a 2d surface in  $\left(m, \rho, H\right)$.  Usually one would parametrize this surface by the coordinates $\left(m, \rho \right)$. However, here it is not very convenient because the constraint  
(\ref{e:H2}) has complicated branches, and this coordinatisation works only locally.\footnote{For the discussion of complications arising due to a similar branching, see e.g.\ \cite{Felder:2002jk}.} In fact the best choice of coordinates here is  $\left(m, H \right)$. This is also very helpful for our purpose of analysing the bouncing solutions.  Indeed it is easy to solve the Friedmann equation with respect to $\rho$ and substitute the result into  Eq. \eqref{e:Hdot} and Eq.~\eqref{e:ddphi}. We have 
\be
  \rho\left(m, H \right)= 3H^2 -3\kappa m H - mK_{m}+K \,. 
\ee 
As we assume standard stable matter such as dust, radiation, etc., the positivity of $\rho$ restricts the region(s) in phase space to which the original physical system can evolve.  Note that curves $\rho\left(m, H \right)=0$ correspond to the 1d phase space describing dynamics of the shift-symmetric scalar field in a universe containing no accompanying matter. Thus these curves are solutions of the equations of motion and the trajectories never cross these boundaries.	   

In shift-symmetric systems, the charge density \eqref{e:n} reduces to  $n\left(m,H\right)=K_{,m} + 3H\kappa$ and is conserved. Having eliminated $\rho$, we can write the dynamics of the universe as the first-order autonomous system:
\begin{align} \label{MHEoM}
&\dot m=\frac{3\kappa\left(1+w\right)\left(K+3H^{2}-mn\right)-3n(2H-\kappa m)}{2n_{m}+3\kappa^{2}} \ ,  \\
&\dot H=-\frac{3\kappa nH+n_{m}\left(\left(1+w\right)\left(K+3H^{2}-mn\right)+nm\right)}{2n_{m}+3\kappa^{2}} \ . 
\end{align} 
Owing to shift-charge conservation, this system is integrable i.e.\ possesses the following first integral 
\be \label{firstIntegral}
\mathcal{I}\left(m,H\right)=\frac{n^{1+w}}{\rho}=\frac{\left(K_{,m}+3H\kappa\right)^{1+w}}{3H^{2}+K-m\left(K_{,m}+3H\kappa\right)}=\text{const}\,. 
\ee
Each trajectory can be parameterised by the value $\mathcal{I}_0$ of the first integral and is given by the solution of $\mathcal{I}\left(m,H\right)=\mathcal{I}_0$.  
Further one can substitute $\dot{m}\left(m,H\right)$ into (\ref{generalSound})
and obtain
\be
c_{\text{s}}^{2}\left(m,H\right)=\frac{f_{2}\left(m\right)H^{2}+f_{1}\left(m\right)H+f_{0}\left(m\right)}{4mD^{2}}\,, \label{soundShift}
\ee
where 
\begin{align*}
 & f_{2}\left(m\right)=6(5+3w)\kappa\kappa_{,m}\,,\\
 & f_{1}\left(m\right)=2\kappa\left(12\kappa^{2}-3m(1+3w)\kappa\kappa_{,m}+8K_{,mm}\right)\,,\\
 & f_{0}\left(m\right)=-3m\kappa^{4}+6\kappa\kappa_{,m}\left((1+w)K-mwK_{,m}\right)+4K_{,m}K_{,mm}+\kappa^{2}\left(6K_{,m}-2mK_{,mm}\right)\,.
\end{align*}
$c_{\text{s}}^{2}\left(m,H\right)=0$ is a quadratic equation with
respect to $H$, so that it appears always possible to chose $\kappa$,
$w$ and $K$ such that the sound speed is positive in the whole region
where $D>0$.

At the bounce, denoted with the subscript $0$, the positive energy density of the fluid is compensated for 
by the negative energy density of the scalar field:\footnote{One could expect that this cancellation might be a substantial source of isocurvature perturbations.} 
\begin{equation}
\rho_{0}=-\mathcal{E}_{0}=\left.\left(K-mK_{,m}\right)\right|_{0}>0\,.\label{PositiveRhoBaunce}
\end{equation}
Further at the bounce  
\begin{equation} \label{IntegralBounce}
\mathcal{I}_{0}=\left. \frac{\left(K_{,m}\right)^{1+w}}{K-mK_{,m}}\right|_0 \,,
\end{equation}
for some $\mathcal{I}_{0}$.%
\footnote{From here one can see that for normal matter, with $w<1$, the trajectories
which can bounce (or crunch) and go to, or appear from $H=\pm\infty$
with finite $m$, build measure zero. Indeed, 

\[
\lim_{H\rightarrow\infty}\mathcal{I}\left(m,H\right)=3^{w}\kappa^{1+w}H^{w-1}\,,
\]
thus $\mathcal{I}\left(m,\infty\right)=3\kappa^{2}$ for $w=1$ and
$\mathcal{I}\left(m,\infty\right)=0$ for normal matter with $w<1$. Thus for these trajectories $\mathcal{I}_{0}=0$ , and  (\ref{IntegralBounce}) generically has a finite countable number of solutions.}

In order to realise a bounce and not a recollapse one has to require that
\begin{equation}
\dot{H}_{0}=-\frac{K_{,mm}\left(\left(1+w\right)K-wmK_{,m}\right)}{2K_{,mm}+3\kappa^{2}}>0\,.\label{dotH>0}
\end{equation}
Taking into account the \textit{no ghost} inequality $2D_{0}=2K_{,mm}+3\kappa^{2}>0$
we arrive at the following two options at $m_{0}$ :
\begin{align}
-\frac{3}{2}\kappa^{2}<K_{,mm}<0 & \quad\text{and\quad}\left(\left(1+w\right)K-wmK_{,m}\right)>0\,,\label{NegativeK''}\\
K_{,mm}>0 & \quad\text{and\quad}\left(\left(1+w\right)K-wmK_{,m}\right)<0\,.\label{PositiveK''}
\end{align}
Negative $K_{,mm}$ at the bounce generically implies that somewhere
in phase space $D<0$. Note that the boundary of $D=0$ is a pressure
singularity \cite{Barrow:2004xh, Deffayet:2010qz, Pujolas:2011he}. Finally
we have to require the absence of gradient instabilities: $c_{\text{s}}^{2}>0$.
At the bounce this implies that
\begin{align}\label{PostiveSoundBounce}
mf_0(m)=&-3m^{2}\kappa^{4}+6m\kappa\kappa_{,m}\left((1+w)K-mwK_{,m}\right)+\\
&\quad +4mK_{,m}K_{,mm}+\kappa^{2}m\left(6K_{,m}-2mK_{,mm}\right)>0\,.\notag 
\end{align}
Finally we would like to stress that we are only considering the stability with respect to high-frequency perturbations. In particular, we have ignored the possible tachyonic masses and related instabilities, e.g.\ Jeans instability.  Note that perturbations $\delta \phi$ and $\delta \rho$ diagonalise the equations for cosmological perturbations in the short-wavelength limit only.
\subsection{The Hot G-Bounce}\label{sec:hotgbounce}

In this section we will present an example from a class of models which bounce in a healthy manner and then proceed to evolve to a phase mimicking radiation domination. We will model the bounce as occurring in the presence of radiation. However, as we will show, the presence of external matter is not actually necessary and its equation of state is largely insignificant. This sort of bouncing trajectory is possible even in a universe containing only the kinetically braided scalar and no external fluid.

In particular, we will analyse the model which, in addition to standard Einstein gravity and an external radiation fluid with equation of state parameter $w=1/3$, contains a minimally coupled kinetically braided scalar with the Lagrangian functions
\begin{equation}\label{hot:lagr}
	K = X - \alpha^2 X^3 = \frac{m^2}{2} - \frac{\alpha^2 m^6}{8} \qquad \text{and} \qquad \kappa = 2\varkappa X = \varkappa m^2\,.
\end{equation}
In the function $K$ above, the term $X^2$ is missing, therefore it appears that the model is fine tuned. However, we have only picked such a model since it is the presence of the $X^3$ term that is the essential component driving the dynamics discussed below and such a simplified choice makes the calculations somewhat more tractable. Adding the $X^2$ term back in---or indeed even higher powers of $X$---does not prevent the model from having largely the same behaviour, provided the coefficients are appropriately chosen. The coefficient of $X^2$ needs to be somewhat smaller than $\alpha$ (but can be of the same order) without much change to our conclusions (see figure~\ref{shiftyfigs} for the phase portrait for a \emph{hot G-bounce} model including an $X^2$ term).

The charge density \eqref{e:n} is 
\begin{equation}
	n = m\left(1 - \frac{3}{4}\alpha^2 m^4 + 3H\varkappa m\right)\,.
\end{equation}
The equations of motion can be obtained from Eqs~\eqref{MHEoM}. It is convenient to rescale time in these equations:
\begin{equation}
	t = \tau\sqrt{\alpha}\,,
\end{equation}
so that the chemical potential $m=\dot\phi$ and the Hubble parameter $H=\dot a / a$ will also correspondingly rescale to become dimensionless,
\begin{equation}\label{hotscaled}
	m=\frac{\mu}{\sqrt{\alpha}} \qquad \text{and} \qquad H = \frac{h}{\sqrt{\alpha}}\,.
\end{equation}

Here, we should also comment on the constant $\alpha$. It has dimension $M_\alpha^{-4}$. Since one would expect this mass scale, $M_\alpha$, to typically be below $\Mpl$, this implies that in Planck units $\alpha$ is some number larger than one. As we will see, the bounce and associated dynamics occur in the region $\mu\sim 1$, $h<1$. This means that by picking a low scale for $\alpha$, we can make the physical scale of the bounce significantly below $\Mpl$ and thus ensure that gravity is under control, away from its quantum regime.

In these new variables, the system~\eqref{MHEoM} takes the form
\begin{align}\label{hot:system}
	h'=&\frac{\left(15 \mu^4-4\right) \left(48 h^2+4 \mu^2+\mu^6\right)-96\beta  h \mu \left(12 h^2+2 \mu^2+\mu^6\right) -144 \beta^2 h^2 \mu^4 }{24 \left(4+24 \beta h \mu  +3 \mu^4 \left(2\beta ^2-5\right)\right)} \ , \\
	\mu'=& \frac{\mu \left[\beta \mu^3 \left(4+\mu^4\right)  -6 h \left(4+\mu^4 \left(2 \beta ^2-3\right)\right)-24 \beta h^2 \mu \right]}{2\left(4+24 \beta h \mu  +3 \mu^4 \left(2 \beta^2-5\right)\right)} \ . \notag
\end{align}
where $'$ denotes differentiation with respect to $\tau$ and where we have defined a rescaled diffusivity parameter
\begin{equation}
	\beta \equiv \frac{\varkappa}{\alpha}\,,
\end{equation}
which will play the role of the sole parameter in this system. This is a quantity of mass dimension one and we will denote as $\beta$  the numerical coefficient of the Planck mass in this ratio:
\begin{equation}
\beta = \left(\frac{M_\alpha}{M_\varkappa} \right)^3 \left( \frac{M_\alpha}{\Mpl} \right)\ .
\end{equation}
where $M_\varkappa$ is the mass scale associated with $\varkappa$.

\paragraph{Stability of Bounce}

Let us now show under which conditions the bounce is stable. As we will demonstrate, there is a range of values of $\beta$ for which a stable bounce can take place.

First, let us evaluate $h'_0$ at the bounce point:
\begin{equation}
	h_0' = \frac{\mu^2\left(\mu^4+4\right)\left( 15 \mu^4 - 4 \right)} {96D_0} \,.
\end{equation}
This is positive provided that $\mu^4 > 4/15$ and the no-ghost condition may also be satisfied
\begin{equation}
	D_0 = 1 + \frac{3}{2} \mu^4 \left( \beta^2 - \frac{5}{2} \right) > 0 \,, 
\end{equation}
from which two conditions arise
\begin{align}
	&\beta^2 \geq \frac{5}{2} \qquad\text{and}\qquad \mu^4 > 0 \,, \\
	\text{or}\quad & \beta^2 < \frac{5}{2} \qquad\text{and}\qquad \mu^4 < \frac{4}{15-6\beta^2}\,.
\end{align}
As we will show, the models with $\beta^2>5/2$ do not bounce stably (see the condition arising from positivity of sound speed, Eq.~\eqref{hot:cspos}), therefore we have the requirement that for a ghost-free bounce
\begin{equation}\label{hot:gfrange}
	\frac{4}{15} < \mu^4 < \frac{4}{15-6\beta^2} \,.
\end{equation}
In this model, $\mu^4=\mu^4_0$ is the minimal value of $\mu^4$ beneath which the energy for the scalar becomes positive at the bounce and would need to be compensated by a negative external energy density: this is the boundary of the dynamically inaccessible region,
\begin{equation}\label{hot:0energy}
	\mathcal{E}_0 = \frac{\mu_0^2}{2}\left(\frac{5}{4}\mu_0^4-1\right)=0 \qquad \Rightarrow \qquad \mu_0^4 = \frac{4}{5} \,.
\end{equation}
The lowest value of $\beta^2$ is such that the maximum ghost-free $\mu^4$ as given by the inequality Eq.~\eqref{hot:gfrange} is also the minimum $\mu^4$ as given by Eq.~\eqref{hot:0energy}, i.e.\
\begin{equation}
	\beta^2 > \beta^2_\text{g} \equiv \frac{5}{3}\,.
\end{equation}

The upper bound on $\beta^2$ comes from considering the sound speed at the bounce. For the \emph{hot G-bounce} model, the positivity condition \eqref{PostiveSoundBounce} reduces to
\begin{equation}\label{hot:poscs}
	\mu^8 \left(\frac{45}{4} + \beta^2 - 3\beta^4 \right) - 2\mu^4 \left(9-4\beta^2\right)+4 > 0\,.
\end{equation}
Again, the condition that there be at least one trajectory which has a stable bounce is equivalent to asking that the bounce be stable when the external energy density vanishes, i.e. $\mu^4 = \mu_0^4$. Substituting this into Eq.~\eqref{hot:poscs} gives a condition on $\beta$:
\begin{equation}\label{hot:cspos}
\beta^2 < \beta^2_\text{c} \equiv \frac{1}{3}\left(7+\sqrt{34}\right) \,.
\end{equation}
For $\beta$ exceeding this upper bound, the sound speed at the bounce is always negative. In summary, stable bounces occur when
\begin{equation}
	1.29 \simeq \beta_\text{g} < |\beta| < \beta_\text{c} \simeq 2.07 \,.
\end{equation}
This condition is independent of the equation of state of the external fluid, provided that $w>-1$. This is the case since we have derived this condition by looking at the limiting trajectory where the external energy density vanishes.

One may then ask whether it is possible to bounce while keeping the perturbations subluminal, $c_\ssl^2 \leq 1$. The answer is of course yes, since for $\beta$ close to $\beta_\text{c}$ there are very few trajectories where sound speed squared is at all positive and for all of them it is close to zero. Let us be more precise: one needs to look at the full sound-speed expression, Eq.~\eqref{soundShift}, which in the case of the \emph{hot G-bounce} reduces to
\begin{equation}
	c_{\ssl, 0}^2 = \frac{16 + 8 \mu^4 (4 \beta^2-9) + 
 \mu^8 (45 + 16 \beta^2 - 12 \beta^4)}{(4 + 
  3 \mu^4 (2 \beta^2-5))^2} \ ,	
\end{equation}
at the bounce point. This can be solved by requiring that a trajectory which bounces when the sound speed is exactly equal to the speed of light is stable to obtain a new limiting value,\footnote{The remaining solutions are not relevant since they fail the other stability tests.}
\begin{equation}
	\beta_1^2 = \frac{49+\sqrt{241}}{24},
\end{equation}
Then trajectories which bounce stably with subluminal sound speed exist provided that
\begin{equation}
	1.64 \simeq \beta_\text{1} < |\beta| < \beta_\text{c} \simeq 2.07 \,.
\end{equation}
One should note that the sound speed typically increases following the bounce and can easily become superluminal. We should stress that this does not result in any causal paradoxes \cite{Babichev:2007dw} even though it does signify that the UV completion of this model would not be Lorentz invariant \cite{Adams:2006sv}.

\paragraph{Further Evolution}

\begin{figure}[t]
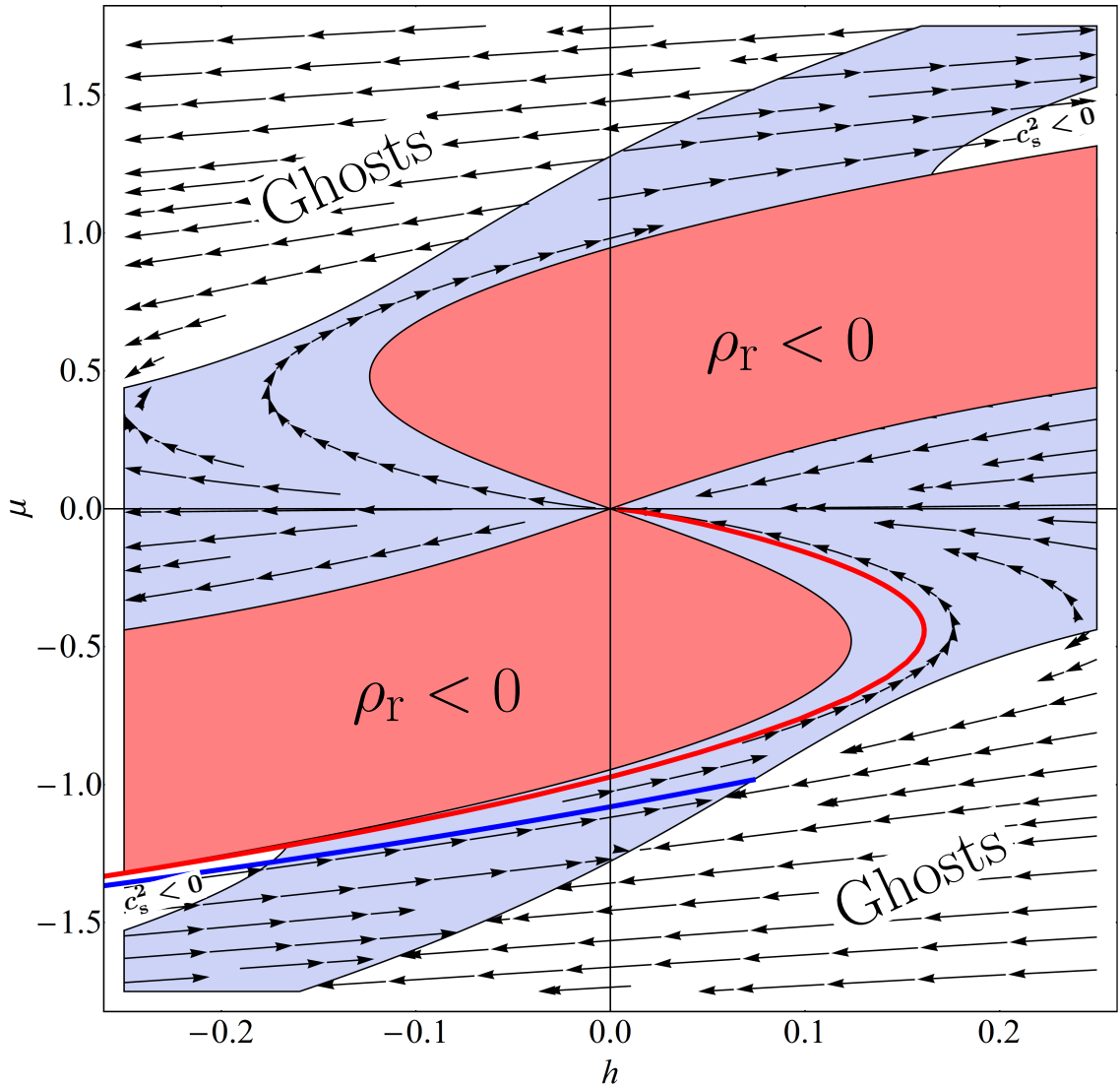
\begin{center}
\begin{lpic}[]{x-x3-w03_b15(15cm)}
\lbl[W]{400,1050,25;{\Huge Ghosts}}
\lbl[W]{1050,250,25;{\Huge Ghosts}}
\lbl[w]{190,225,15;{\small $\boldsymbol{c_\text{\textbf{s}}^2<0}$}}
\lbl[w]{1200,1100,15;{\small $\boldsymbol{c_\text{\textbf{s}}^2<0}$}}

\lbl[]{500,450;{\Huge $\rho_\text{r}<0$}}
\lbl[]{900,850;{\Huge $\rho_\text{r}<0$}}

\end{lpic}
\caption{ \label{f:hot} Plot of the phase space for the \emph{hot G-bounce model} \eqref{hot:lagr} in the presence of external radiation fluid ($w=1/3$), $\mu=\sqrt{\alpha}\dot\phi$, $h=\sqrt{\alpha}H$, $\beta=\varkappa/\alpha=1.5$. The solution is fully under control only in the blue region: pink regions are dynamically inaccessible, while the white regions suffer from instabilities (e.g.\ negative sound speed squared, $c_\text{s}^2 <0$, or ghosts, $D<0$.) In red, we have marked an example trajectory: it enters the healthy region from one where $c_\text{s}^2<0$, bounces, reaches a maximum Hubble parameter and then proceeds toward the origin. The equation of state of the scalar approaches $w_X=1$, so eventually it redshifts away, leaving only the accompanying radiation fluid. In blue, we have marked a nearby trajectory: here the Hubble parameter does not turn around, but the trajectory proceeds to a pressure singularity ($D=0$) where $\dot{H}$ diverges (but both the scale factor and Hubble parameter remain finite). Only a narrow range of bouncing trajectories arrive safely in the late-time radiation-domination era; the rest end in singularities.}
\end{center}\end{figure}

Having established that it is possible to bounce stably in the \emph{hot G-bounce} model, let us now turn to analysing the phase space numerically. We refer the reader to figure~\ref{f:hot}, where we have plotted the interesting part of the phase space. The striking feature of the \emph{hot G-bounce} model is that there exist trajectories which after bouncing reach a maximum Hubble parameter and turn around to evolve toward the origin of the phase space---Minkowski spacetime.

%
%

For any value of $h$ there are limiting values of $\mu$, for which the external energy density vanishes (the boundaries of the red regions in figure~\ref{f:hot}) beyond which the system cannot evolve. The dynamically accessible region then is carved out by the inequality
\begin{equation}\label{hot:dynacc}
	3h^2 > \alpha\mathcal{E} = \frac{1}{2}\mu^2 - \frac{5}{8}\mu^6 + 3h\beta \mu^2 \ .
\end{equation}
In the vicinity of the origin, $\mu, h \ll 1$ so \eqref{hot:dynacc} reduces to 
\begin{equation}
	h \geq  -\frac{\mu}{\sqrt{6}} + \mathcal{O}(\mu^3) \,,
\end{equation} 
where we have picked the trajectories in the lower-right quadrant of figure~\ref{f:hot}. 

The first type of trajectory we will consider is one corresponding to a universe with no accompanying external fluid: the evolution proceeds exactly along the boundary of the dynamically inaccessible pink region in this case. Taking the solution along the boundary close to the origin as $h\simeq -m/\sqrt{6}$, we obtain as an approximation for the system \eqref{hot:system}
\begin{align}
 	h' &= -3h^2 +\Ocal(h^4)  \,,\\
 	\mu'&= \sqrt{\frac{3}{2}}\mu^2 + \Ocal(\mu^3) \,. \notag
\end{align}
We can solve this approximate system to obtain
\begin{align}
 	h &\simeq \frac{1}{3\tau}\,,\qquad\qquad \tau \rightarrow \infty \,,\\
 	\mu &\simeq -\frac{2}{\sqrt{6}\tau} \,, \notag
\end{align}
which is an expanding universe comprising a fluid with a constant equation of state $w_X=1$. This is unsurprising, since at late times, when $\mu\ll 1$, only the leading-order term $X$ in the Lagrangian function $K$ is relevant. Thus the evolution is just that of a kinetic-energy-dominated canonical scalar field.

In order to analyse the approach to the origin of the other trajectories, we turn to the first integral of the equations of motion, Eq.~\eqref{firstIntegral}. In the vicinity of the origin the leading-order contribution to the first integral is
\begin{equation}
	\Ical \simeq \frac{(-\mu)^{4/3}}{3h^2}\,.
\end{equation}
This implies that close to the origin
\begin{equation}	
	\mu \simeq -(3\Ical)^{3/4} h^{3/2} \,,
\end{equation}
since this is the only solution which conserves the first integral as the origin is approached. Again, approximating the system \eqref{hot:system} on such a solution, we obtain
\begin{align}
 	h' &= -2h^2 - \frac{\sqrt{3}}{2}\Ical^{3/2} h^3 + \Ocal(h^4) \,, \label{hot:h} \\
 	\mu' &= \sqrt{\frac{3}{\Ical}}(-\mu)^{5/3} + \Ocal(\mu^{10/3})\,. \label{hot:mu}
\end{align}
Eq.~\eqref{hot:mu} can be solved easily to obtain
\begin{equation}
	\mu \simeq -\left(\frac{3\Ical}{4}\right)^{3/4} \tau^{-3/2} \,, \qquad\qquad \tau\rightarrow\infty\,.
\end{equation}
We can also solve Eq.~\eqref{hot:h} provided we change the time variable to the scale factor, $a$. We then obtain the solution
\begin{equation}
	h = \left(Ca^2 - \sqrt{\frac{3\Ical^3}{4}}\right)^{-1}\,, \qquad\qquad a\rightarrow\infty \,.
\end{equation}
with $C$ some constant of integration.  By bringing this to the more user-friendly form of the Friedmann equation, we obtain
\begin{equation}
	h^2 \simeq \frac{1}{C^2a^4} + \frac{\sqrt{3\Ical^3}}{C^3a^6}\,, \qquad\qquad a\rightarrow \infty \,,
\end{equation}
which represents a radiation-domination phase with a small correction coming from a fluid with $w=1$.

We can also calculate the properties for the perturbations of the scalar in the vicinity of the origin on the above background solution. We obtain
\begin{align}
	D &\simeq 1 - 6\beta (3\Ical)^{3/4} h^{5/2}\,, \qquad\qquad h\rightarrow 0 \,,\\
	c^2_\ssl &\simeq 1+ 8\beta(3\Ical)^{3/4} h^{5/2} \,. \notag
\end{align}
This is again what we would have naively expected assuming that all the higher-order terms in $K$ and $G$ become irrelevant and we are effectively dealing with a kinetic-energy-dominated canonical scalar field in the presence of radiation.

\paragraph{Perturbations and Possible Inflationary Stage}

In this paper, we have not dealt with the perturbation spectrum and how it could match up to the observational limits. Models featuring an NEC-violating stage in the early universe presented to date produce the wrong spectrum unless furnished with an additional scalar field which acts as a curvaton \cite{Creminelli:2010ba,Levasseur:2011mw, Qiu:2011cy}. We discuss these models further in \S~\ref{sec:cb}.

In the case of the \emph{hot G-bounce} model presented here, we also have some stumbling blocks. Firstly, the change in the scale factor during the collapsing pre-bounce stage is actually very small: only a few percent when the moment of crossing $c_\text{s}^2=0$ is taken as the comparison point. This is the case for the \emph{hot G-bounce} model and all the models presented in figures~\ref{shiftyfigs}~and~\ref{shiftyfigs2}.\footnote{The pressure singularities, $D=0$, occur at a finite scale factor and Hubble parameter. The only divergent quantities are $\dot{H}$ and sound speed.}  A stable collapsing universe per se is not difficult to construct---it has always been the bounce itself that was problematic and we have solved this problem here. However, we have been unable to graft a long-lived collapsing phase onto our bouncing trajectories in a simple way. We remain hopeful that given enough effort this will turn out to be possible.

Even without the collapsing phase one could imagine producing the spectrum of perturbations in a way akin to tachyacoustic inflation \cite{ArmendarizPicon:2003ht, ArmendarizPicon:2006if, Piao:2006ja,Bessada:2009ns,Liu:2011ns}: As the trajectory approaches the pressure singularity, the sound speed becomes very high and then decreases as the system moves away. A decreasing sound speed provides for an acoustic horizon that shrinks (in comoving coordinates) and under the right circumstances is able to solve the horizon problem even for a non-inflating background. A sufficiently large causal horizon could be set up during evolution along a trajectory which comes very close to the singularity.\footnote{One would have to worry about remaining within the weak-gravity regime and away from strong-coupling. Picking the right mass scale for $\alpha$ to achieve this may not be possible for trajectories too close to the singularity.} However, in the \emph{hot G-bounce} model, trajectories which pass close to the pressure singularity have large and rapidly-varying slow roll parameter $\dot{H}/H^2$ and sound speed which would give a naive two-point function that is not scale-invariant. At this point, one would need to ensure that the standard Bunch-Davies vacuum is even established and to what extent the standard calculation of perturbation freeze-out carries through. This may not be the case for such a rapidly changing acoustic horizon \cite{Kinney:2007ag}. Whether it would be possible to generate an appropriate spectrum through such a mechanism warrants further investigation and we leave it to future work.

For the moment, as a simple fix, we would like to point out that nothing prevents us from adding in a positive cosmological constant to the \emph{hot G-bounce} model. This has the effect of making the final state of the evolution a de Sitter rather than a Minkowski space-time.  In figure~\ref{shiftyfigs}, we show that bouncing trajectories of the type that we have discussed above are in fact still present in the phase space, provided that the cosmological constant is not excessively large.

We could then imagine that this cosmological constant is in fact a result of the existence of a very flat potential for the kinetically braided scalar $\phi$, flat enough to be effectively constant until significantly after the bounce.  We can then replicate a standard inflationary mechanism following the bounce. Just as in the standard inflationary scenario, a flat enough potential will give the correct perturbation spectrum following reheating. One may then ask whether it would be possible to have a sufficiently long stage of inflation to explain the perturbation spectrum at large scales, but retain some of the perturbations generated during the bounce at sufficiently large scales; for possible observational consequences see e.g. \cite{Piao:2003zm}.

As such, the \emph{hot G-bounce} model would not provide an alternative for inflation, but would be a possible extension of the inflationary mechanism into the past. We do not, of course, have full control of the trajectory at all times into the past: all the bouncing trajectories originate in regions with gradient instabilities. This is similar to the situation with inflation where the trajectories are also incomplete to the past, implying the existence of an initial singularity \cite{Borde:2001nh}. However, in this model, the energy scale of the crossover from regions of instability is in fact arbitrarily low (controlled by the mass scale in $\alpha$), perhaps allowing us to avoid having to deal with quantum gravity.

\subsection{Other Shift-Symmetric G-bounce Models}\label{sec:othershift}

\begin{figure}[t]
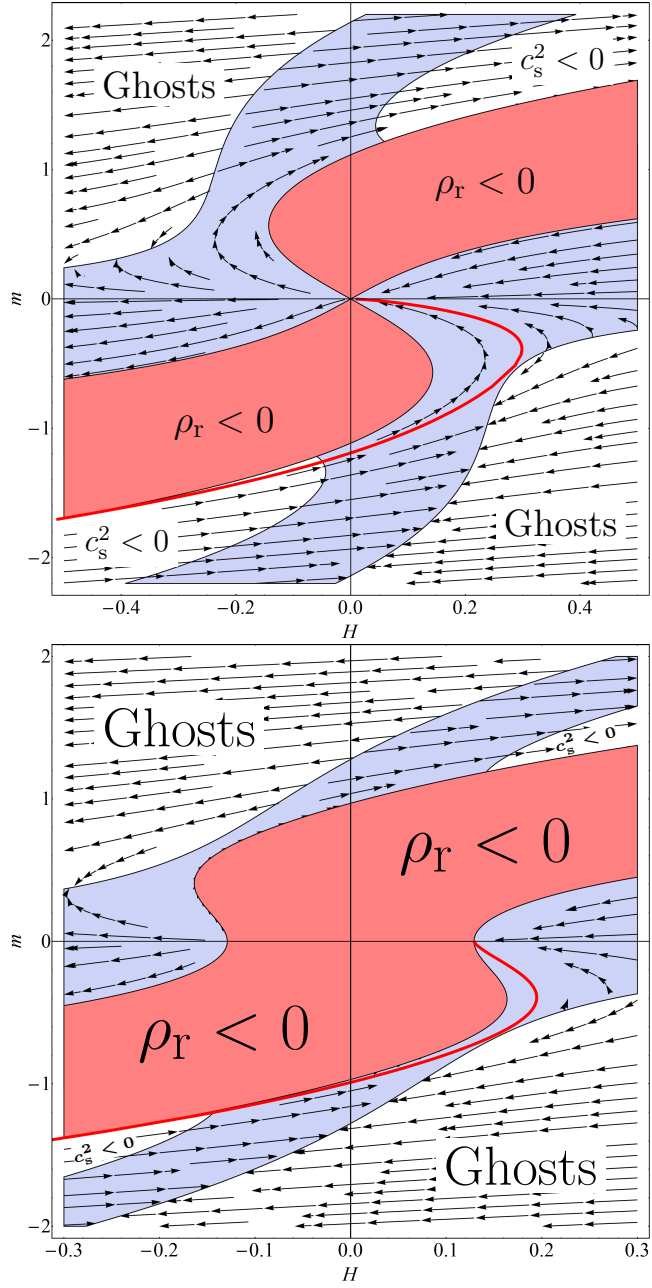
\begin{center}

\begin{lpic}[]{x+05x2-x3-w03_b15(8.5cm)}
\lbl[W]{200,700;{\Large Ghosts}}
\lbl[W]{700,150;{\Large Ghosts}}
\lbl[W]{702,730;{\large $c_\text{s}^2<0$}}
\lbl[W]{160,125;{\large $c_\text{s}^2<0$}}
\lbl[]{275,275;{\Large $\rho_\text{r}<0$}}
\lbl[]{600,575;{\Large $\rho_\text{r}<0$}}
\end{lpic}
\begin{lpic}[]{x-x3-dS-w03_b15(8.5cm)}
\lbl[W]{225,700;{\huge Ghosts}}
\lbl[W]{650,150;{\huge Ghosts}}
\lbl[w]{725,685,15;{\tiny $\boldsymbol{c_\text{\textbf{s}}^2<0}$}}
\lbl[w]{125,170,15;{\tiny $\boldsymbol{c_\text{\textbf{s}}^2<0}$}}
\lbl[]{275,315;{\Huge $\rho_\text{r}<0$}}
\lbl[]{600,550;{\Huge $\rho_\text{r}<0$}}
\end{lpic}

\caption{\label{shiftyfigs} A selection of phase-space portraits for shift-symmetric \emph{G-bounce} models with external matter. The classical evolution presented here is only under control in the blue regions: the pink colouring represents dynamically inaccessible regions and the white regions have unstable perturbations (either negative sound speed squared or ghosts). \\ \textbf{Top Panel:} \emph{Hot G-bounce} model with the $X^2$ term restored: $K=X+X^2/2 -X^3$, $\kappa=2\beta X$, $\beta=1.5$ in the presence of an accompanying radiation fluid. Restoring the $X^2$ term does not substantially change the phase space presented in figure~\ref{f:hot}, provided its coefficient is somewhat smaller than one. \\
\textbf{Bottom Panel:} \emph{Hot G-bounce} model with an additional positive cosmological constant: $K=-\Lambda+X-X^3$, $\kappa = 2\beta X$, $\Lambda=0.05$, $\beta=1.5$, accompanying radiation fluid. Despite the addition of $\Lambda$, bouncing trajectories still exist, but now proceed to a late-time de Sitter attractor, providing an inflationary stage following the bounce.}
\end{center}\end{figure}

In figures~\ref{shiftyfigs}~and~\ref{shiftyfigs2} we have provided the phase-space diagrams for a number of shift-symmetric \emph{G-bounce} models, demonstrating the wide range of trajectories possible in this class. In all the diagrams, only the blue regions are under full control: we have marked as pink the dynamically inaccessible regions and as white the regions plagued by instabilities. The supplied bestiary of models is by no means exhaustive: the bounces are simple to achieve and the stable violation of the NEC opens up a whole set of alternative expansion histories for the early universe.


\begin{figure}[t]\begin{center}

\begin{lpic}[]{mx-x3-w1_b17(8.5cm)}
\lbl[W]{250,600;{\huge Ghosts}}
\lbl[W]{650,250;{\huge Ghosts}}
\lbl[]{225,250;{\Large $\rho_\text{stiff}<0$}}
\lbl[]{650,615;{\Large $\rho_\text{stiff}<0$}}
\lbl[w]{700,758,20;{\small $\boldsymbol{c_\text{\textbf{s}}^2<0}$}}
\lbl[w]{150,100,20;{\small $\boldsymbol{c_\text{\textbf{s}}^2<0}$}}
\end{lpic}
\begin{lpic}[]{x-x3-AdS-w03_b17-g001(8.5cm)}
\lbl[W]{175,525;{\Large Ghosts}}
\lbl[W]{700,325;{\Large Ghosts}}
\lbl[W]{700,625;{\Large $c_\text{s}^2<0$}}
\lbl[W]{175,225;{\Large $c_\text{s}^2<0$}}
\lbl[]{250,350;{\Large $\rho_\text{r}<0$}}
\lbl[]{650,512;{\Large $\rho_\text{r}<0$}}
\end{lpic}

\caption{\label{shiftyfigs2} A selection of phase-space portraits for shift-symmetric \emph{G-bounce} models with external matter. Color coding as in figure~\ref{shiftyfigs}. \\
 \textbf{Top Panel:} \emph{G-bounce} with destabilised Minkowski: $K=-X-X^3$, $\kappa=2\beta X$, $\beta=1.7$ in the presence of stiff matter, $w=1$. In this type of models, the trajectories generically begin in a pressure singularity and then cross $c_\text{s}^2=0$ (red) after bouncing, or vice-versa (blue). These models contain de Sitter attractors/repellers similar to those described in Ref.~\cite{Deffayet:2010qz} and trajectories end or begin there. \\
  \textbf{Bottom Panel:} bounce \& recollapse model: $K= -\Lambda + X - X^3$, $\kappa=2\beta X + \gamma X^2$, negative $\Lambda = -0.05$, $\beta = 1.7$, $\gamma=-0.04$ with radiation, $w=1/3$. The red trajectory begins in a singularity at large $H$. The expansion slows down until the universe begins to recollapse. The crunching is prevented, however, and the universe bounces. The trajectory then crosses $c_\text{s}^2=0$ eventually evolving to an unstable de-Sitter attractor. The blue trajectory is the time reverse: first a bounce, then a recollapse.}
\end{center}\end{figure}
\paragraph{Hierarchy of Sufficient Conditions} As we have argued above, a stable bounce is realised when all the inequalities \eqref{PositiveRhoBaunce}, \eqref{dotH>0}, \eqref{PostiveSoundBounce} and \eqref{e:D} hold.  
Now one can ask how simple is it to satisfy all these inequalities?  Are our examples above, see figures (\ref{f:hot}), (\ref{shiftyfigs}) and (\ref{shiftyfigs2}),  very special / fine-tuned? Here we provide simple \emph{sufficient} conditions for a healthy bounce.  
Namely the following conditions guarantee a healthy bounce: 
\begin{itemize}
\item $\kappa\kappa_{,X}>0$;
\item matter is normal $w\geq -1$. Apart from this requirement, the models are indifferent as  to the equation of state of external matter;
\item the following hierarchy is satisfied at the bounce
\begin{equation}
K>mK_{,m}>\frac{1}{2}\left(m\kappa\right)^{2}>0>\frac{1}{2}m^{2}K_{,mm}>-\frac{3}{4}\left(m\kappa\right)^{2}\,.\label{hierarchy}
\end{equation}
\end{itemize}

Indeed the first inequality from the left guarantees \eqref{PositiveRhoBaunce}
and $K+w\left(K-mK_{,m}\right)>0$. Further the last two inequalities
yield $D>0$ therefore we get \eqref{NegativeK''}, i.e. $\dot{H}_{0}>0$.
Now let us see whether one can have a positive sound speed. Using
$m\kappa\kappa_{m}=2X\kappa\kappa_{X}>0$ we obtain 
\[
mf_{0}\left(m\right)>-3m^{2}\kappa^{4}+2\left(2mK_{,m}-\left(m\kappa\right)^{2}\right)K_{,mm}+6\kappa^{2}mK_{,m}\,,
\]
then using the second and third inequalities from \eqref{hierarchy}
we obtain
\[
mf_{0}\left(m\right)>-3m^{2}\kappa^{4}-3\left(2mK_{,m}-\left(m\kappa\right)^{2}\right)\kappa^{2}+6\kappa^{2}mK_{,m}=0\,.
\]
This hierarchy \eqref{hierarchy} of inequalities is sufficient but \emph{not necessary} for a stable bounce. However, having two free functions $\kappa$ and $K$ it is easy to chose them to satisfy  \eqref{hierarchy} for some range of $m$. In that case this system will bounce in a healthy fashion in this chosen range of $m$ for any type of external matter with $w\geq -1$.  In particular it seems to be possible to choose $G$ and $K$ to satisfy the hierarchy for \emph{all} $m$.  Note that it is also easy to construct a theory which allows for a stable bounce but violates the hierarchy, see e.g. the system on the top panel of figure (\ref{shiftyfigs2}) where the violation is manifest.  
These sufficient conditions are rather intuitive and can help for a future engineering of bouncing cosmologies.  

\clearpage

\section{Bounces in Models with Negligible External Matter}\label{s:nomatter}

Models in which the external matter can be considered negligible in the vicinity of the bounce provide another category of bouncing cosmologies with a significantly simplified, two-dimensional phase space. We will discuss such models in this section and specialise to the discussion of the so-called \emph{conformal Galileon} model \cite{Nicolis:2008in, Creminelli:2010ba, Levasseur:2011mw}.

The coordinates for the phase space in the class are provided by the pair $(\phi, m)$. However, the Friedmann equation \eqref{e:H2} can be used, at least in principle \emph{locally},  to eliminate the field value $\phi$, by substituting it with a function of $m$ and $H$,
\begin{equation}\label{e:phisol}
	\phi = \phi_*(m,H)\,.
\end{equation}
This gives a phase space similar to the one obtained in section \S\ref{sec:siffsymm}  with coordinates $(m,H)$, with the advantage again that the bounce position in the phase space is very explicit. On the other hand, the Lagrangian may have complicated dependence on $\phi$ and therefore the solution Eq.~\eqref{e:phisol} may have many branches, and it may not even be possible to obtain a closed form version of it. Nonetheless, in the example considered below, such a procedure does bear fruit.

The evolution in this phase space is described by
\begin{align}\label{e:solfevol}
	\dot{m} &= \frac{-3n \left(2H - \kappa m \right)- 2mn_{,\phi} +2K_{,\phi}}{2D}\,, \\
	\dot{H} &= \frac{\kappa\left(K_{,\phi}-mn_{,\phi}\right) - n\left(mn_{,m}+m\kappa_{,\phi} + 3\kappa H \right)}{2D}\,. \notag
\end{align}
with $D$ defined in Eq.~\eqref{e:D} and any dependence on the field value $\phi$ eliminated through Eq.~\eqref{e:phisol}.
The above results are too general to proceed further. We restrict our attention to an example already known from the literature: the \emph{conformal Galileon}.

\subsection{Conformal G-Bounce Model} \label{sec:cb}

The action for the so-called \emph{conformal Galileon} model is \cite{Creminelli:2010ba,Nicolis:2008in}:\footnote{For further development see \cite{Levasseur:2011mw} and \cite{Qiu:2011cy}.}
\begin{equation}
S_{\phi}=\int\mbox{d}^{4} \! x\sqrt{-g}\left[- 2 f^2 e^{2 \phi} X+ \frac{2 f^3}{\Lambda^3} X^{2}+ \frac{2 f^3}{\Lambda^3}  X\Box\phi\right] \,,
\label{cg:action}
\end{equation}
where $\Lambda$ and $f$ are constants with mass dimension one. We will consider this model as being minimally coupled to gravity. The conformal Galileon can be rewritten in tems of our notation as 
\begin{align}
	K &= -f^2 e^{2\phi}m^2 + \frac{f^3}{2\Lambda^3}m^4 \,,\\
	\kappa &= \frac{2f^3}{\Lambda^3} m^2 \,.
\end{align}

We proceed with the analysis of background cosmology in this model by applying the method described in \S\ref{s:nomatter}: we can eliminate the scalar field value through the Friedmann equation,
\begin{equation}\label{cg:dynnotac}
	e^{2\phi} = \frac{3f^3 m^4 + 12 H f^3 m^3 -6\Lambda^3 H^2} {2f^2 \Lambda^3 m^2}\,.
\end{equation}
This quantity must be strictly positive. This implies that wherever the above function is negative will be a region of the phase space which is dynamically inaccessible.

Equations of motion for $H$ and $m$ can now be easily derived using their general equivalents \eqref{e:solfevol}.  It is helpful to rescale the time variable which enters both $m$ and $H$,\footnote{As opposed to the rescaled variables in the \emph{hot G-bounce} model defined in Eqs~\eqref{hotscaled}, here both $h$ and $\mu$ remain dimensionful (and are presented in Planck units).}
\[\label{cg:time rescale}
t=\left(\frac{f}{\Lambda}\right)^{3/2}\tau\,.
\]
We then obtain a rescaled chemical potential, $\mu$, and a rescaled Hubble parameter $h$,
\begin{align}\label{cg:rescaled variables}
m=\mu\left(\frac{\Lambda}{f}\right)^{3/2} \qquad \text{and} \qquad   H=h\left(\frac{\Lambda}{f}\right)^{3/2} \,,
\end{align}
and we will denote differentiation with respect to this new time variable with a prime, $(\ )'\equiv\ds/\ds \tau$. This procedure greatly simplifies the equations of motion and yields a set of equations which are \emph{free of parameters}:

\begin{align}
\label{cg:MuEvolv}
\mu'&=\mu\frac{\mu^{3}\left[\left(12h^{2}+5h\mu+\mu^{2}\right)-\mu^{3}\left(6h+\mu\right)\right]-2h^{2}(3h+\mu)}{\mu^{4}\left(2\mu^{2}+1\right)+2h^{2}} \,, \\
 h'&=\frac{\mu^{3}\left[\mu^{3}\left(12h^{2}+16h\mu+3\mu^{2}\right)-8h^{2}\mu\right]-12h^{4}}{2\left(\mu^{4}\left(2\mu^{2}+1\right)+2h^{2}\right)} \,. \notag
\end{align}
%
This observation implies that the phase space topology is independent of the choice of the parameters $f$ and $\Lambda$ and only the physical scales corresponding to various trajectories will differ as $f,\Lambda$ are changed. 

We are now able to write down the conditions for stability Eqs~\eqref{e:D} and \eqref{generalSound} in the conformal Galileon model in terms of the rescaled variables:
\begin{align}\label{cg stability}
	D&=\frac{6h^2}{\mu^2} + 3\mu^2 + 6\mu^4 >0\ , \\
	c_\text{s}^2 &= 3(\mu^2 D)^{-2} \left[ 12h^{4} -32h^{3}\mu^{3}+4h^{2}\mu^{4} \left(14\mu^{2}-1\right)+\right.\label{cg cs2} \\
	 &\qquad\qquad\qquad+\left.16h\mu^{7}\left(1-2\mu^{2}\right) + \mu^{8}\left(3-4\mu^{2} \left(2+\mu^{2}\right)\right)\right] > 0 \,.\notag
\end{align}
It can immediately be seen that $D>0$ at all points. Close to the origin, values of $h$ are constrained by the requirement to keep Eq.~\eqref{cg:dynnotac} positive, which gives us $h < \mu^2/\sqrt{2}$. This implies that for all the trajectories in the dynamically accessible region, $D=0$ at the origin. For all trajectories leaving or entering the origin $\ds h/\ds\mu=0$ and using l'H\^{o}pital's rule one can see that $D\rightarrow 0$ on those trajectories.

Let us now deal with the scales present in this model. Firstly, the physical quantities (in Planck units) are related to the rescaled quantities defined in Eqs~\eqref{cg:rescaled variables} through relations of the type 
\begin{align}\label{cg:physical}
	H^2 = \left(\frac{\Lambda}{f}\right)^3 h^2 \ , \ \qquad
	\dot{H} = \left(\frac{\Lambda}{f}\right)^3 h' \ .
\end{align}
This implies that introducing a hierarchy $\Lambda\ll f$ will reduce the physical scales in the problem, allowing larger regions of phase space to retain curvatures significantly below order $\Mpl^2$.

Secondly, as discussed in Ref.~\cite[Eq.~(17)]{Creminelli:2010ba}, there exists a strong-coupling scale, which in Minkowski can be estimated to be
\begin{equation}\label{eq:strong}
	\Lambda_\text{Strong}^2 \sim \Lambda^2 e^{2\phi} \ .
\end{equation} 
This implies that in the part of phase space immediately surrounding the dynamically inaccessible region (the boundary of which is the surface $e^{2\phi}=0$, see Eq.~\eqref{cg:dynnotac}), the strong-coupling scale is always very low and approaches zero. This means that it is the trajectories which approach the boundary of the dynamically inaccessible region that will cross into a strong-coupling regime for the perturbations and not necessarily just a trajectory which has a large value of $h$. It should be noted, however, that a proper calculation of the strong-coupling scale taking into account the background solution has not been performed, thus one should not at this stage treat the strong-coupling problems calculated in this way as definite.

\paragraph{Conformal Bounce Conditions}

Specialising the result Eq.~\eqref{cg:MuEvolv} to the bounce point, $h=0$, we obtain 
\begin{equation}\label{cg hdot}
	h'_0 = \frac{3\mu_0^4}{2+4\mu_0^2} > 0\,.
\end{equation}
Since in the conformal Galileon model this quantity is always positive, the bounce point can only be crossed from collapse to expansion, if at all. Any trajectory that bounces will not proceed to recollapse, even if it may approach $h=0$. 

The stability at the bounce point is determined by the sign of the quantity Eq.~\eqref{cg cs2}, which reduces to the condition
\begin{equation}
	4\mu_0^4 + 8\mu_0^2 -3 < 0 \ .
\end{equation}
This can be solved, giving
\begin{equation}
	0 < \abs{\mu_0} < \frac{\sqrt{7}}{2}-1 \approx 0.568 \,.
\end{equation}
Substituting these limits into Eq.~\eqref{cg hdot} gives us the constraint that during stable bounces
\begin{equation}	
	0 < h_0' < \frac{7\sqrt{7}-17}{16} \approx 0.095 \,,
\end{equation} 
or, equivalently in physical units
\begin{equation}
	0 < \dot{H}_0 < 10^{-1} \left(\frac{\Lambda}{f}\right)^3\Mpl^2 \,.
\end{equation}
The above shows that obtaining a stable bounce within the conformal Galileon model is in principle possible, as was noted in \cite{Creminelli:2010ba} and demonstrated in detail in \cite{Qiu:2011cy}. 

\begin{figure}[t]\begin{center}
\begin{overpic}[width=15cm]{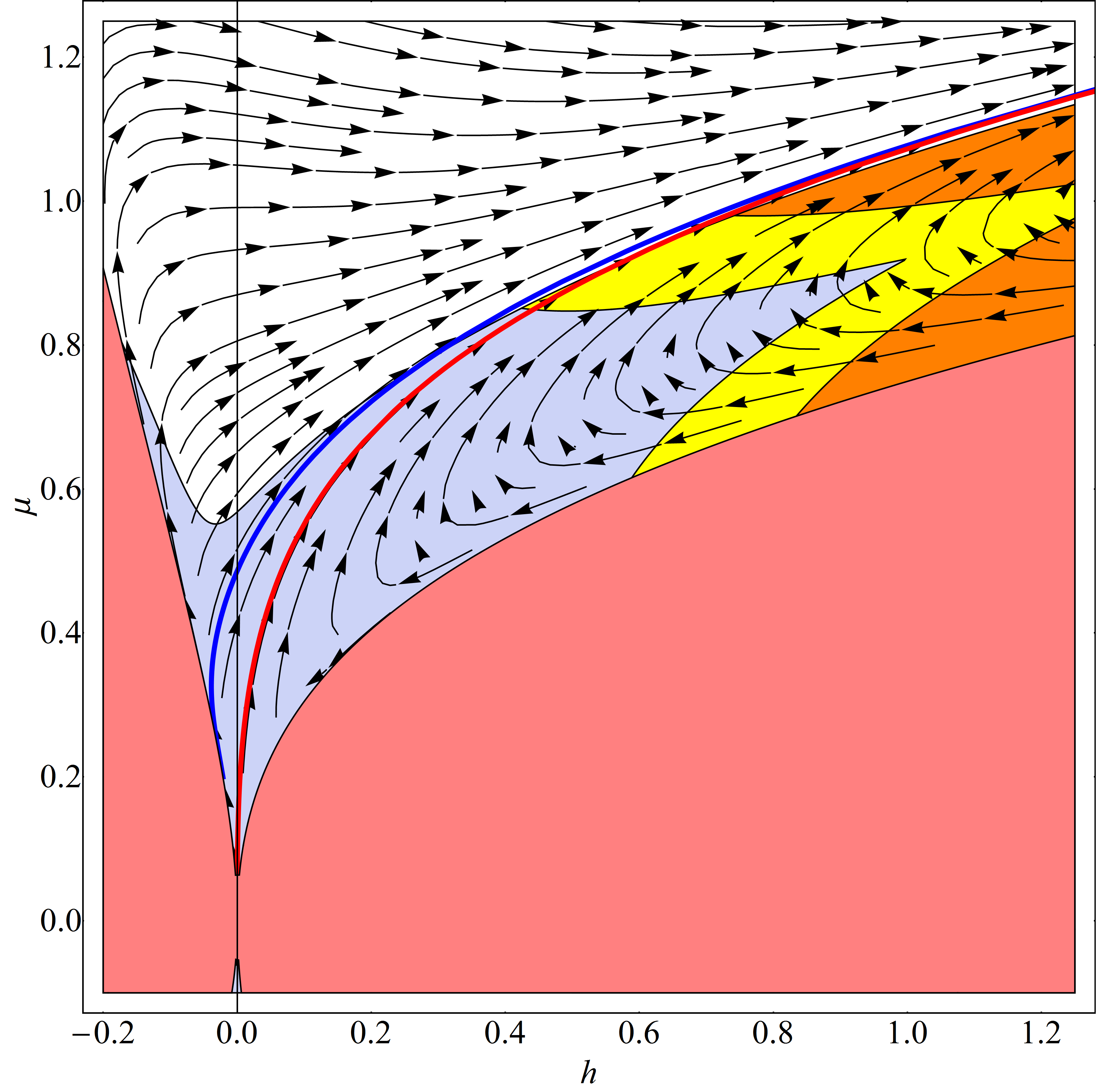}
\put(225,42){\fboxsep 0pt \fboxrule 1pt%
 \fbox{\includegraphics[width=6.7cm]{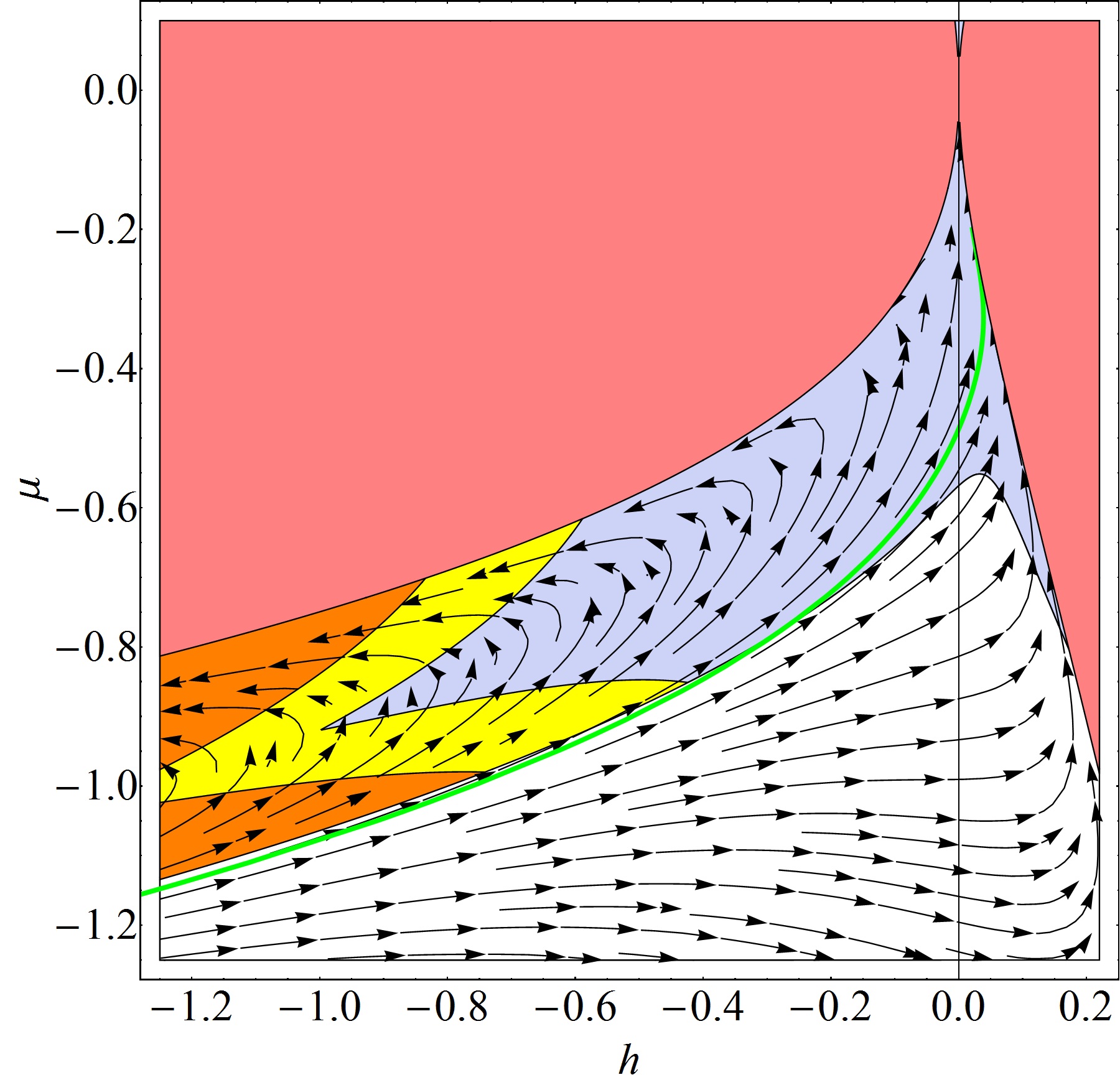}}}
\put(100,360){\colorbox{white}{\fboxsep 1pt \fboxrule 1pt \fbox{\Huge $c_\ssl^2<0$}}}
\put(70,70){\Large Dynamically inaccessible}
\put(110,90){\Huge $e^{2\phi}<0$}
\put(320,80){\colorbox{white}{\fboxsep 0.5pt \fboxrule 0.5pt \fbox{\small $\boldsymbol{c_\text{\textbf{s}}^2<0}$}}}
\put(260,180){\small Dynamically inaccessible}
\put(295,165){$e^{2\phi}<0$}
\end{overpic}

\caption{\label{confphase.eps} Phase portrait for the conformal Galileon model, Eq.~\eqref{cg:action} in rescaled coordinates $h=(f/\Lambda)^{3/2}H$, $\mu=(f/\Lambda)^{3/2}\dot \phi$. The main figure shows the stable region where $\mu>0$. The inset depicts the time-reversed region with $\mu<0$. The solution is under control fully only in the light blue regions: pink corresponds to dynamically inaccessible regions, white---to negative sound speed squared. Yellow and orange are regions where curvature is transplanckian for $(f/\Lambda)^3=1,2$, respectively. The blue line is a typical healthy bouncing trajectory (presented in  \cite{Qiu:2011cy}): it originates from a region where the theory is strongly coupled, but the background solution evolves as a collapsing radiation-dominated cosmology; the universe then bounces in a healthy region and then the trajectory very rapidly crosses into the region where $c_{\text{s}}^{2}<0$  and the classical solution should not be trusted. The red trajectory is the \emph{Galilean Genesis} trajectory \cite{Creminelli:2010ba}: it begins in the vicinity of the Minkowski origin; the universe is always expanding and eventually the trajectory crosses the line $c_{\text{s}}=0$ around $h=0.6$; depending on the choice of parameters this happens either before or after the curvatures become transplanckian. Both the trajectories merge to an attractor which evolves toward a Big Rip singularity. In the inset in green, we have marked a trajectory time-reversed with respect to the blue discussed above: this one begins in a (collapsing) Big Rip singularity, at some point crosses into a region of positive sound speed squared, bounces and then proceeds to expand in a radiation-domination-like phase which is also strongly coupled.}
\end{center}\end{figure}

We now turn to a global analysis of the phase space in order to understand the past and future of the trajectories which exhibit a bounce. We refer the reader to figure~\ref{confphase.eps} and its caption for a description of the dynamics.


\paragraph{Evolution of Trajectories}

Firstly, we comment on the behaviour of the evolution close to the origin of the phase plot which is a highly degenerate singular point. From physical considerations it is clear that the origin is a fixed point. There are no other finite fixed points for this system. In particular, in the dynamically allowed region close to the origin, $|h| < \mu^2/\sqrt{2}+ \Ocal(\mu^3)$. We can then find an equation describing trajectories in the vicinity of the origin, $h=h_*(\mu)$, by solving order by order for the coefficients of a Taylor series approximating the solution of 
\begin{equation}\label{cg:trajs}
 	\dbd{h}{\mu}  = \frac{\mu^{3}\left[\mu^{3}\left(12h^{2}+16h\mu+3\mu^{2}\right)-8h^{2}\mu\right]-12h^{4}}{2\mu\left(\mu^{3}\left[\left(12h^{2}+5h\mu+\mu^{2}\right)-\mu^{3}\left(6h+\mu\right)\right]-2h^{2}(3h+\mu)\right)} \,.
\end{equation}
This allows us to obtain two solutions. The first is the equation for the boundaries of the dynamically inaccessible region which, as has already been stated, are also solutions. Taking $h<0$:
\begin{equation}\label{cg:raddom}
	h_{*\text{b}} = -\frac{\mu^2}{\sqrt{2}} + \mu^3 - \frac{\mu^4}{\sqrt{2}} + \cdots = \mu^3 - \frac{\mu^2}{\sqrt{2}} \sqrt{1+2\mu^2} \,.
\end{equation}
On this trajectory, the system of equations \eqref{cg:MuEvolv} near the origin reduces to 
\begin{align}
	\mu' &= \frac{\mu^3}{\sqrt{2}} + \Ocal(\mu^4) \,,\\
	h' &= -\mu^4 + \Ocal(\mu^5) \,. \notag
\end{align}

If we induce the Minkowski space to collapse, then the universe evolves on the approximate solution
\begin{align}
	\mu &\simeq \frac{1}{(-\tau\sqrt{2})^{1/2}}\,,\qquad\qquad \tau\rightarrow -\infty\\
	h &\simeq  \frac{1}{2\tau}\,, \notag
\end{align}
which corresponds to a collapsing radiation-domination era. On this trajectory the perturbations have the following properties:
\begin{align}
	D &= 6\mu^2 + \Ocal(\mu^3)\,, \qquad\qquad \mu\rightarrow 0 \\
	c_\ssl^2 &= \frac{1}{3} - \frac{\mu^2}{3} + \Ocal(\mu^3) \,.
\end{align}

A trajectory such as this one has been marked as blue on figure~\ref{confphase.eps} and was studied in \cite{Qiu:2011cy} where the above evolution corresponds to the initial pre-bounce stage of the solution.

The second solution to \eqref{cg:trajs} corresponds to a separatrix between the bouncing trajectories in the left-upper quadrant of figure~\ref{confphase.eps} and those trajectories which never collapse, but begin and end in a Big Rip and fill the top-right quadrant of the figure:
\begin{equation}
	h_{*\text{g}} = \frac{\mu^3}{2} + \frac{3}{10}\mu^5 + \frac{3}{35}\mu^7 + \Ocal(\mu^9)\,.
\end{equation}
This trajectory is in fact the \emph{Galilean Genesis} trajectory of \cite{Creminelli:2010ba}. We can again approximate the system of equations \eqref{cg:MuEvolv} on this trajectory 
\begin{align}\label{cg:orig}
	\mu' &= \mu^2 + \Ocal(\mu^4)\,,\\
	h'  &= \frac{3}{2} \mu^4 + \mathcal{O}(\mu^6) \,. \notag 
\end{align}
If we start slightly away from Minkowski along the \emph{Genesis} trajectory, then we obtain an evolving solution which is always expanding. Eqs~\eqref{cg:orig} can be solved:
\begin{align}
	\mu &\simeq -\frac{1}{\tau}\,, \qquad\qquad \tau\rightarrow -\infty\,,\\
	h &\simeq -\frac{1}{2\tau^3}\,. \notag
\end{align}
\clearpage
\noindent
On these trajectories in the vicinity of the origin then
\begin{align}
	D &= 3\mu^2 + \Ocal(\mu^4)\,, \qquad\qquad \mu\rightarrow 0 \\
	c_\ssl^2 &= 1 - \frac{16\mu^2}{3} + \Ocal(\mu^4) \,.
\end{align}
This matches the asymptotics obtained in \cite{Creminelli:2010ba}. We have plotted this trajectory as the red line in figure~\ref{confphase.eps}.

If we evolve the \emph{Galilean Genesis} trajectory forward, it will cross into the region with negative sound speed squared. This occurs at $h_\text{c}\simeq 0.6$. Depending on the value of the Lagrangian parameters $f, \Lambda$, this may happen before or after gravity becomes strong, but is inevitable for all the trajectories. Indeed, all trajectories which have bounced in the past also eventually approach and cross the boundary $c_{\text{s}}^2=0$.\footnote{In fact all the bouncing trajectories enter the region where perturbations are out of control at an even smaller value of $h$, $h<0.6$}  

On figure~\ref{confphase.eps} it is hard to see that the transition through this boundary $c_{\text{s}}^2=0$ occurs. This curve is given by the solution of the polynomial equation $F\left(\mu,h\right) \equiv \left(\mu^2 D\right)^2 c_{\text{s}}^2=0$. If some trajectories (of nonzero measure) do not cross $c_{\text{s}}^2=0$ then there should be an interval of $h$ where these trajectories approach this boundary with $\ds\mu / \ds h=-F_{,h} / F_{,\mu}$, i.e.\ system of equations $F\left(\mu,h\right) =0$ and $\ds\mu / \ds h=-F_{,h} / F_{,\mu}$ should have a continuum of real solutions. As it is easy to check this is not the case and this polynomial system has 22 isolated complex roots. Thus all trajectories do evolve across $c_{\text{s}}^2=0$, because this curve is neither a trajectory nor a singularity of the system \eqref{cg:MuEvolv}. 

Importantly, this \emph{Galilean Genesis} trajectory does not evolve toward the region where the perturbations are strongly coupled as naively defined in Eq.~\eqref{eq:strong}, but in fact evolves \emph{away}. Thus it is either strong gravity or gradient instabilities that will signify that the solution is failing, unless the effective field theory leaves its regime of validity.\footnote{See the discussion in \cite{Nicolis:2009qm} regarding the rather subtle question of the appropriate cut-off for the effective field theory.} In fact at the beginning  
of the \emph{Galilean Genesis} trajectory, where $h=0$ and $\mu=0$,  the naive strong coupling scale $\Lambda_{\text{Strong}}$, given by Eq. \eqref{eq:strong}, is exactly zero. One would expect that this indicates that the scalar field is infinitely strongly coupled there.  
\\
Moreover, when one integrates the \emph{Galilean Genesis} trajectory numerically, one finds that the system effectively spends all of its time in the vicinity of the origin. Once $h$ begins to pick up, the evolution is extremely fast and the Big Rip is reached in a short time. This means the time during which the system is in the region where $h$ is significantly different from zero and yet there are no gradient instabilities is actually very brief. Hence the scale factor increases only by approximately 50\% between the beginning of the trajectory in the vicinity of the origin and the trajectory's leaving the stable region, i.e.\ only about 0.4 e-folds of expansion are under control for the perturbations. In the \emph{Galilean Genesis} scenario effectively all the expansion occurs in the region where gradient instabilities are present in the degree of freedom driving the expansion. 

Also in the bouncing trajectories of the type \eqref{cg:raddom} the amount of expansion between the bounce and the crossing into $c_\text{s}^2<0$ is quite small. However, here all the perturbations are set up during the collapsing phase, which can effectively be arbitrarily long. 

One should be concerned, however, since the naive strong coupling scale \eqref{eq:strong} is in fact extremely low in the whole region where the contraction mimics radiation-domination and the perturbations are generated. The proper strong-coupling scale for a cosmological background should be calculated in order to judge this with any certainly. This is outside of the scope of this work. Despite the fact that the observable perturbations are driven by a curvaton, a separate  scalar field, the predictivity of the calculation for curvaton perturbations is also uncertain. The curvaton by design is coupled explicitly to the conformal galileon, thus if the galileon perturbations are strongly coupled, the curvaton perturbations will also not be under control.

\section{Conclusions}\label{sec:con}

In this paper, we have demonstrated that stable bouncing cosmologies are generic and simple to achieve in models containing non-canonical scalar fields with \emph{Kinetic Gravity Braiding}. We have constructed models that can realise a transition from contraction to expansion in a flat Friedmann universe stably, without leaving the weak-gravity regime. The ingredients can---but do not have to---include external matter, which will generically blue shift as the universe contracts. 

The conditions for the bounce to be healthy are that the perturbations remain ghost free and that the evolution is free of gradient instabilities. Indeed, we have derived a set of sufficient conditions on the form of the Lagrangian of the theory which will ensure that the evolution at the turnaround point itself is stable. As we have shown, constructing Lagrangians which contain healthy bouncing trajectories is not difficult. 

On the other hand, we have not succeeded in constructing a model where the whole expansion history, including the remote past and remote future, is under control. We have found that, generically, all trajectories which bounce usually have some kind of pathology. A typical bouncing trajectory begins or ends in a Big Rip or pressure singularity. At best, the trajectories cross the line of vanishing sound speed, corresponding to a singularity in the acoustic metric, which can appear in non-canonical theories \cite{Babichev:2007dw}. Since quantum fluctuations are normalised by the sound speed, they will grow without bound as the trajectory approaches the singularity. However, this is a singularity in the scalar sector, the energy scale of which is essentially arbitrary. Thus, these trajectories avoid being transplackian and as such do not necessarily involve strong-gravitational effects.

We have succeeded in finding a category of models which, while starting from such an acoustic singularity, bounce stably and then evolve to a healthy future, the precise nature of which depends on the accompanying fluid. The late-time solutions can correspond, for example, to a hot Big-Bang---when the external fluid is radiation---or to a standard inflationary stage---if the scalar field is furnished with a very flat potential which breaks shift-symmetry weakly.

Despite this success, the bouncing models we have proposed do not resolve the problem of the initial singularity of the universe, similarly to inflation \cite{Borde:2001nh}. One could say that we have shifted back the beginning of the universe's clock to a collapsing, pre-inflationary stage.
However, we remain optimistic that one should be able to eventually resolve this issue. The models we have presented give an idea of what a solution could look like: a model with a fixed point inside a stable region could have trajectories which are healthy throughout their complete history and never evolve to a strong-gravity regime. The ease with which one can construct models featuring stable bounces gives hope that some additional complication (for example, interactions of the braided scalar and other degrees of freedom) will result in a realistic bouncing model of the universe.


\acknowledgments

It is a pleasure to thank Luca Amendola, Eugeny Babichev, Ram Brustein, Yi-Fu Cai, Riccardo Catena, Jarah Evslin, Slava Mukhanov and Taotao Qiu for helpful conversations. A.V.\ would like to thank the organizers of PiTP 2011 program at the IAS, Princeton for kind hospitality. I.S.\ would like to thank Deutsche Bahn for the comfort and efficiency of its ICE service, which facilitated the final stages of this project. The work of D.A.E.\ is supported in part by the Cosmology Initiative at Arizona State University.  I.S.\ is supported by the DFG through TRR33 ``The Dark Universe''. A.V.\ is supported in part by grant FQXi-MGB-1016 from the Foundational Questions Institute (FQXi) through Theiss Research. 

\bibliographystyle{utphys}
\addcontentsline{toc}{section}{References}
\bibliography{bounce}{}

\end{document}